\documentclass{PoS}
\usepackage{amsmath,amsfonts,euscript,amssymb}
\def\be{\begin{equation}}
\def\ee{\end{equation}}
\newcommand{\bea}{\begin{eqnarray}}
\newcommand{\eea}{\end{eqnarray}}

\title{{Origin of masses in the Early Universe}}

\ShortTitle{Origin of masses in the Early Universe}

\author{\speaker{Victor Pervushin}%
         \\
        JINR\\
        E-mail: \email{pervush@theor.jinr.ru}}

\author{
    \\ Andrej~Arbuzov,  Alexander~Cherny, Vadim~Shilin,
        \\JINR, Dubna,
\\
Rashid Nazmitdinov, \\
JINR,~Dubna~and~Department~de~F{\'\i}sica,~Universitat~de~les~Illes~Balears,~Palma~de~Mallorca,~Spain,
\\Alexander Pavlov, \\
Russian State Agrarian University -- Moscow Timiryazev Agricultural Academy, Moscow, 127550, Russia,\\
Konstantin~Pichugin,\\
Kirensky Institute of Physics, Krasnoyarsk, 660036 Russia
\\Alexander Zakharov, \\Institute of Theoretical and Experimental Physics, Moscow, 117259, Russia.
 }

\abstract{New model is suggested, where the Casimir mechanism is the source of masses and conformal symmetry breaking at the Planck epoch in the beginning of the Universe. The mechanism is the Casimir energy and associated condensate, which are resulted from the vacuum postulate and normal ordering of the conformal invariant Hamiltonian with respect to the quantum elementary field operators.  It is shown that the  Casimir top-quark condensate specifies the value of the Higgs particle mass without involving the Higgs tachyon mass, which is put equal to zero. The Casimir mechanism  yields another value of the coupling constant for the self-interaction of scalar field than the standard model does.
 }

\FullConference{XXII International Baldin Seminar on High Energy Physics Problems \\
                 September 15-20, 2014\\
                 JINR, Dubna, Russia}

\begin{document}

\section{Introduction}

The basis of modern studies on the conformal symmetry breaking at the present day epoch is the high-energy dimensional  transmutation parameter \cite{cj:1971,cw:1973}, founded upon the Euclidean space formulation of quantum field theory (QFT) and the renormgroup analysis. However, in the paper \cite{ZHETF:72}, Kirzhints developed a new approach to the origin of masses in the Standard electro-weak Model in the Early Universe. Kirzhnits kept the zero harmonic of the Higgs scalar field in finite space\footnote{In agreement with the Einstein  cosmological principle of 1917, this zero harmonic is  the average of a scalar field over the large volume \cite{einstein:1917}.} and suggested the model of resurrection of the conformal symmetry at the electro-weak epoch in the beginning of the Universe.

In the present paper, we develop the Kirzhnits idea and consider the origin of masses and the conformal symmetry breaking via the scalar field zero harmonic within the Minkowskian  space-time.
The Minkowskian  space-time formulation, amending the Kirzhnits theory, allows us to introduce the vacuum postulate and the normal ordering of the conformal invariant action in the  Early Universe.

The second difference from the Kirzhnits paper \cite{ZHETF:72} is the cosmological model of
the \textit{empty Universe} \cite{Behnke:2002,Zakharov:2010nfempty-6,PervushinZakharov2012-6}
without any massive dust and radiation \cite{Arbuzov:2010fz5-5,grg12chapter5}. This model suggests a dominance of  the Casimir vacuum energy
$\langle 0|{\bf H}|0\rangle$, which can be calculated after corresponding regularization from the normal ordering of fields in the Hamiltonian ${\bf H}$ mentioned above. As is shown in the persent paper, the normal ordering leads also to the Casimir condensates of \textit{elementary fields},
defined as the vacuum expectation values of the product of two elementary fields. These Casimir condensates can be sources of the
 masses in the Early Universe.

 The third difference from the Kirzhnits paper \cite{ZHETF:72}
is using the conformal time-space variables (coordinates, fields,  and other magnitudes)
instead of the world variables.
The conformal magnitudes simplify the QFT formulation.
However, these magnitudes require  initial data and boundary conditions that physically differ from the
world initial data and boundary conditions at the beginning of the Universe. In limit of small factor $a \to 0$
we have  long distance conformal intervals $dr=d\eta$, zero conformal masses $m_{\rm conf}(a)\to 0$,
and a cold empty Universe with a small conformal temperature.
These conformal variables and coordinates
  can be treated as the real observational quantities
\cite{Behnke:2002,Zakharov:2010nfempty-6,PervushinZakharov2012-6,Arbuzov:2010fz5-5,grg12chapter5,
  Barbashov:2005hu5-5,pervushin_pavlov_book}.
In this case, a source of the red shift ${1+z}$ is the mass $m_0/(1+z)$,
instead of the photon length $\lambda_0/(1+z)$ as it is accepted in the standard cosmology.
In fact, observers of the modern Supernovae distances (SNe Ia data) \cite{Riess2001-1}
   measure dimensionless number
of the ratio of a photon length $\lambda_0$ and its units associated with the size of atom $m^{-1}_0$.
The SNe Ia data admit both the possibilities \cite{Behnke:2002,Zakharov:2010nfempty-6,PervushinZakharov2012-6}.
In this paper we consider the origin of masses and condensates at the Planck epoch in the framework of
the conformal cosmological model of the cold and empty Universe.

\section {The Electro-Weak Epoch Resurrection}

We shall also use the second Kirzhnits idea
  about the initial data of the symmetry restoration
  at the electro-weak epoch \cite{ZHETF:72}.
  Using the   experience of estimation of the corresponding
   z-factors at the instants of recombination and chemical evolution,
 one can suppose that the corresponding z-factor at the  resurrection epoch can be
  at the instant. Namely, when the world CMB  temperature $T_{\rm w}(a_{\rm EW})=\dfrac{T_{\rm CMB}}{a_{\rm EW}}$
   coincides with the present day
  electro-weak scale, identified with the value of t-quark mass $M_t=174$ GeV in the Standard cosmology variables
\begin{eqnarray}\label{e-w-1}
M_{t0} = \dfrac{T_{\rm CMB}}{a_{\rm EW}}\simeq 174 {\rm GeV}, ~~~~
a^{-1}_{\rm EW} \equiv 1+z_{\rm EW}\simeq 0.32 \times 10^{15}.
\end{eqnarray}
 In terms of the conformal \textit{variables},
 the E-W scale equation (\ref{e-w-1}) transforms to the form
 \be\label{e-w-11}
 M_t(a)\big|_{a=a_{\rm EW}}\equiv M_{t0}\times a_{\rm EW}=T_{\rm CMB}.
 \ee
 This equation describes a small particle mass in the Minkowskian
 space-time $ds^2=(d\eta)^2-(dx)^2$
  at constant temperature $T_{\rm CMB}$.

In terms of the conformal \textit{variables}, one can
  suppose that in the limit of massless particles, before the time of resurrection, the Universe
 was empty but not stable due to a strong  interaction of the Higgs field
 with the cosmological scale factor. The primordial Higgs scalar particle decay becomes a source
 of $N_\gamma=(1+z_{\rm EW})^{6}$  CMB photons \cite{grg12chapter5}. 
At that time the empty Universe  was filled in by the Casimir vacuum energy of all fields.
 The energy density of this vacuum is a single \textit{external}  input parameter that violates  the
 conformal symmetry of the action. The origin of this \textit{external} parameter is due to a
 finite volume of the Universe limited by its horizon.
If at the Beginning, the Universe was quantum, one can apply the Planck's least action postulate.
In this case, values of the action of the Universe are \textit{quantized}.
Planck's least action postulate
determines the initial value of the cosmological  factor $z_{\rm  Pl}\sim z_{\rm EW}$
and yields the hierarchy  of
cosmological scales, according to their conformal weights weights $w$ and dimensions $d$ in natural units $\hbar=c=1$\cite{grg12chapter5,pervushin_pavlov_book}
\bea\label{h-set-1}
H_{\rm Hubble}&\simeq& H_I\cdot {(1+z_{\rm I})}^{-2} ~~~~~~~ {d=1,w=2}\\\label{r-set-1}
R^{-1}_{\odot}&\simeq& H_I\cdot {(1+z_{\rm I})}^{-1}  ~~~~~~~{d=1,w=1}\\\label{t-set-1}
T_{\rm CMB}&=&H_I \cdot {(1+z_{\rm I})}^{0} ~~~~~~~{d=1,w=0}\\\label{m-set-1}
M_{t0}&\simeq&H_I \cdot {(1+z_{\rm I})}^{1}~~~~~~~ {d=1,w=-1}\\\label{p-set-1}
M_{\rm Planck}&\simeq& H_I\cdot {(1+z_{\rm I})^{2}}~~~~~~{d=1,w=-2}.
\eea
If $z_I=z_{\rm EW}$, one knows the present day values of
the Hubble parameter (\ref{h-set-1}), the inverse size of
Celestial system (\ref{r-set-1}),
the CMB temperature (\ref{t-set-1}), electro-weak scale (\ref{m-set-1}), and the Planck mass (\ref{p-set-1}), correspondingly.
This hierarchy is in an agreement with a cosmological evolution of these fundamental magnitudes $F_{w,d}$.
These magnitudes are
 defined by means of the primordial conformal Hubble parameter $H_I$, their conformal weights $w$, and a dimension $d$
 \be\label{cf-2}
F_{w,d}(z)=\gamma_{F}H^d_I\left(\dfrac{1+z}{1+z_I}\right)^{w}\big|_{z=z_I}=\gamma_{F}H^d_I;~~~~
F_{w,d}(0)=\gamma_{F}H^d_I\cdot (1+z_I)^{-w},
\ee
where $\gamma_{F}$ is of order of a unit.
At the E-W epoch, all these values  are approximately
equal to the initial value of the
Hubble parameter at the Planck epoch $z_I=z_{\rm EW}\simeq z_{\rm Planck}$.
This fact is a phenomenological support of our model of the cold empty Universe,
where the modern Supernovae distances (SNe Ia data) \cite{Riess2001-1}
can be explained by the vacuum energy dominance \cite{Behnke:2002,Zakharov:2010nfempty-6,PervushinZakharov2012-6}.

\section{Conformal Version of the Minimal Standard Model}

 Now, in the framework of the condensate mechanism of conformal symmetry breaking,
 we consider the conformal version of SM (CSM) \cite{cmcsb}.
 In the beginning, in the CSM Lagrangian  one keeps only the scalar field potential term.
 It gives rise to the four-type and the largest mass t-quark -- Higgs field interactions
\be\label{L_int}
 L_{\mathrm{int}} = - \frac{\lambda^2}{8}\phi^4 - g_t \phi~\bar{t}t,
 \ee
where {$g_t=1/\sqrt{2}$}. 
The normal ordering of a fermion pair
$f \bar f=:f \bar f:+ \langle f \bar f \rangle$
yields the condensate density of the fermion field $\langle f\bar f \rangle$ in the  Yukawa interaction term
in Eq.(\ref{L_int}). The t-quark condensate
\be\label{V_cond}
{V_{\rm cond}(\phi)=\dfrac{\lambda^2}{8}\phi^4 - g_t \phi <t\bar t>}
\ee
supersedes the phenomenological negative square mass term in the Higgs potential,
as a consequence of the normal ordering. At the present day $z=0$ we have a set of equations:
\bea\nonumber
V_{\rm cond}(\phi)&=&\frac{\lambda^2}{8}\phi^4-g_t \phi\langle t\bar t\rangle\\\nonumber
\left.\frac{dV_{\rm cond}(\phi)}{d\phi}\right|_{\phi=v_{\rm ext}}&=&{v_{\rm ext}^3}\frac{\lambda^2}{2} - {g_t\langle t\bar t\,\rangle}
=0\\\nonumber
\frac{d^2V_{\rm cond}(\phi)}{d\phi^2}&=&m^2_{h } =\frac{3\lambda^2}{2}v_{\rm ext}^2,
\eea
These equations, the definition $M_t=g_tv_{\rm ext}$,~$g_t\simeq 1/\sqrt{2}$, and a present day experimental
value of the Higgs particle mass $m_{h}=v_{\rm ext}/2$ yield
a value of the t-quark condensate  and the interaction constant:
\be
{\dfrac{\langle t\bar t\rangle}{M_t^3}=\dfrac{\lambda^2}{2g_t^4}\simeq \dfrac{1}{3}}\,,\quad
\lambda^2=\dfrac{1}{6}.
\ee
This ratio of t-condensate value and the cubbed mass is close to the one
  at the Planck epoch, when they are determined via the primordial Hubble parameter $H_I$
$$\dfrac{\langle t\bar t\rangle}{M_t^3}=\dfrac{\langle t\bar t\rangle_I}{H_I^3}\simeq \dfrac{1}{3}$$
as it was calculated  in  the book \cite{pervushin_pavlov_book}) (see pp. 323-325) in according to
the equation (\ref{cf-2}).

\section{Conclusion}

 We suggested 
to induce a spontaneous conformal symmetry breaking in the Standard Model by means of the quantum anomalies.
This enables us to avoid the problem of the regularization of the divergent
tadpole loop integrals in the Coleman-Weinberg mechanism of a dimensional transmutation,
by relating them to condensate values. These values could be hopefully extracted from
experimental observations.
The top quark condensate can supersede the tachyon-like mass term in the Higgs potential.
The suggested mechanism allows to establish relations between condensates and masses, including
the Higgs boson one.
In a sense, we suggest a simple bootstrap between the Higgs and top
fields (and their condensates).
We consider the Higgs boson to be an elementary particle,
without introduction of any additional interaction beyond the SM, contrary to
various technicolor models.
After the spontaneous symmetry breaking in the tree level
Lagrangian, the difference from the SM appears only in the value of the
Higgs boson self-coupling $\lambda$. The latter hardly can be extracted from the LHC
data, but it will be certainly measured at a future linear $e^+e^-$ collider.


\end{document}